\documentclass[preprint,12pt,fleqn]{elsarticle}
\usepackage{amssymb}
\usepackage{amsfonts}
\usepackage[english]{babel}
\usepackage{euscript}
\usepackage{bbm}
\usepackage{xfrac}

\usepackage{color}
\usepackage{mathrsfs}
\usepackage{calrsfs}
\usepackage{amsbsy}

\newcommand{\fl}{\hspace*{-\mathindent}}
\newcommand{\textfrac}[2]{\textstyle{\frac{#1}{#2}}}

\newcommand{\eqref}[1]{(\ref{#1})}

\newcommand{\E}{\EuScript{E}}

\newcounter{example_counter}

\newcounter{definition_counter}
\newcounter{theorem_counter}
\newcommand{\thm}{\addtocounter{theorem_counter}{1}{\sc THEOREM \arabic{theorem_counter}}.\,\,\,}

\newcounter{remark_counter}
\newcommand{\uu}{\pmb{u}}
\newcommand{\qq}{\pmb{q}}
\newcommand{\pp}{\pmb{p}}
\newcommand{\ee}{\pmb{E}}

\newcommand{\crl}{\mathrm{curl}\,}
\newcommand{\ccrl}{\mathrm{curl^2}\,}

\begin{document}

\begin{frontmatter}

\title{A Lax representation, symmetries, and conservation laws  for the  three-dimensional  Euler--Helmholtz equations}

\author{Oleg I. Morozov}
\ead{oimorozov@gmail.com}
\address{Trapeznikov Institute of Control Sciences,
65 Profsoyuznaya Street,  
\\ 
Moscow 117997, Russia}

\begin{abstract}

We consider the three-dimensional Euler--Helmholtz equations for an inviscid incompressible fluid. Under the Poincar{\'{e}} lemma assumption, the incompressibility condition is resolved by introducing a vector potential, leading to a vorticity-type reformulation of the system. The main result is the construction of a Lax representation for the obtained system. 
We compute the Lie algebra of point symmetries and find the zeroth-order cosymmetries together with the associated local conservation laws. Using the Lax representation, we derive a shadow of cosymmetry and find nonlocal conservation laws 
using the construction of the  canonical conservation law on the Whitney sum of the tangent and the cotangent coverings. 
Finally, we establish a Bäcklund transformation between the  tangent and the cotangent coverings and describe its action on symmetries and cosymmetries. 

\end{abstract}


\end{frontmatter}



\section{Introduction}

In this paper, we consider  the three-dimensional Euler--Helmhotz equations for  the motion of an incompressible inviscid fluid expressed in the vortex form. These equations are one of the cornerstones of fluid dynamics. 
Due to its fundamental importance, they have been the subject of extensive research and continues to be a focal point for both theoretical and applied studies, see  \cite{ArnoldKhesin} and references therein.

We assume that the underlying domain satisfies the conditions of the Poincar{\'{e}} lemma. 
This allows us to eliminate the divergence-free condition \eqref{div_u} by introducing a vector potential $\pmb{u}$ such that $\pmb{U}=\crl \pmb{u}$. 
A central result of this paper is the construction of a novel Lax representation for the obtained system \eqref{main_eq}. 
The presence of a Lax pair is widely regarded as the key feature of integrability (see, e.g., \cite{WE,Zakharov82,RogersShadwick1982,NovikovManakovPitaevskiyZakharov1984,Konopelchenko1987,AblowitzClarkson1991,%
MatveevSalle1991,BacklundDarboux2001} 
and references therein). From the geometric perspective of differential equations, Lax representations are naturally formulated in the language of nonlocal symmetries and differential coverings, providing a unified framework for the study of integrable models \cite{KrasilshchikVinogradov1984,KrasilshchikVinogradov1989,VK1999}. 

The Lax representation for the two-dimensional Euler equation in the vorticity form was found in \cite{Li2001}. Two Lax representations for the three-dimensional Euler--Helmholtz equations were presented in \cite{LiYurov2003}. The result of \cite{Li2001} was generalized by applying the technique of twisted extensions of Lie symmetry algebras from \cite{Morozov2022} in the paper 
\cite{Morozov2024a}, where we have derived a family of Lax representations  for the 2D  Euler equation. The construction of \cite{Morozov2024a} can be extended  to 
several {\sc PDE}s describing various hydrodynamics models 
\cite{Morozov2023b,Morozov2023c,Morozov2024b,Morozov2025,Morozov2026b}.
In this  paper, we  generalize the results of \cite{Morozov} and present the Lax representation \eqref{the_covering}  for  the three-dimensional Euler--Helmholtz equations \eqref{main_eq}.

Further, we derive the generators of the algebra of the nontrivial  infinitesimal point symmetries of \eqref{main_eq} and describe the structure of this Lie algebra. We also find the  nontrivial zeroth-order cosymmetries of the system and construct the corresponding local conservation laws. 

Then, following the idea of \cite{Morozov2026}, we find a shadow of a cosymmetry in the obtained covering \eqref{the_covering} and, combining this with the construction of the  canonical conservation law \cite{KrasilshchikVerbovetskyVitolo2017}, obtain  nonlocal conservation laws for \eqref{main_eq}. 

Finally, we establish a Bäcklund transformation between the tangent and the cotangent coverings of \eqref{main_eq}. We analyze the action of this transformation, as well as its inverse, on point symmetries and zeroth-order cosymmetries. In particular, 
the action of the inverse Bäcklund transformation on two symmetries  is expressed in terms of the pseudopotentials of the covering \eqref{the_covering}. 


\section{Preliminaries and notation}

The presentation in this section closely follows
\cite{KrasilshchikVinogradov1989,VK1999,KrasilshchikVerbovetsky2011}.

Let $\pi \colon \mathbb{R}^n \times \mathbb{R}^m \rightarrow \mathbb{R}^n$,
$\pi \colon (x^1, \dots, x^n, u^1, \dots, u^m) \mapsto (x^1, \dots, x^n)$, be a trivial bundle, and
$J^\infty(\pi)$ be the bundle of its jets of infinite order. The local coordinates on $J^\infty(\pi)$ are
$(x^i,u_I^\alpha)$, where $I=(i_1, \dots, i_n)$ are multi-indices, and for every local section
$f \colon \mathbb{R}^n \rightarrow \mathbb{R}^n \times \mathbb{R}^m$ of $\pi$ the corresponding infinite jet
$j_\infty(f)$ is a section $j_\infty(f) \colon \mathbb{R}^n \rightarrow J^\infty(\pi)$ such that
$u_I^\alpha(j_\infty(f)) =\displaystyle{\frac{\partial ^{\vert I\vert} f^\alpha}{\partial x^I}}
=\displaystyle{\frac{\partial ^{i_1+\dots+i_n} f^\alpha}{(\partial x^1)^{i_1}\dots (\partial x^n)^{i_n}}}$.

The  vector fields
\[
D_{x^k} = \frac{\partial}{\partial x^k} +\sum \limits_{\alpha=1}^{m} \sum \limits_{\vert I\vert \ge 0}
u^\alpha_{I+1_{k}}\,\frac{\partial}{\partial u_I^\alpha},
\qquad  k \in \{1,\dots,n\},
\]
$(i_1,\dots, i_k,\dots, i_n)+1_k = (i_1,\dots, i_k+1,\dots, i_n)$,  are called {\it total derivatives}.
They com\-mu\-te everywhere on $J^\infty(\pi)$.

A system of {\sc pde}s $F^r(x^i,u^\alpha_I) = 0$, $r \in \{1,\dots, \varrho\}$, of order $s \ge 1$ with $\vert I\vert \le s$ defines the submanifold $\E=\{(x^i,u^\alpha_I)\in J^\infty(\pi)\,\,\vert\,\,D_K(F^r(x^i,u^\alpha_I))=0,\,\,\vert K\vert\ge 0\}$
in $J^\infty(\pi)$.

Let the linear space $\EuScript{W}$ be either $\mathbb{R}^N$ for some $N \ge 1$ or  $\mathbb{R}^\infty$ endowed with  local co\-or\-di\-na\-tes $w^a$, $a \in \{1, \dots , N\}$ or  $a \in  \mathbb{N}$, respectively. A {\it differential covering} of $\E$ is  a trivial bundle $\varpi \colon J^\infty(\pi) \times \EuScript{W} \rightarrow J^\infty(\pi)$ equipped with 
{\it extended total derivatives}
\[
\widetilde{D}_{x^k} = D_{x^k} + \sum \limits_{a}
T^a_k(x^i,u^\alpha_I,w^b)\,\frac{\partial }{\partial w^a}
\]
such that $[\widetilde{D}_{x^i}, \widetilde{D}_{x^j}]=0$ for all $i \not = j$ if 
$(x^i,u^\alpha_I) \in \E$. 
Define the partial derivatives of $w^a$ by  $w^s_{x^k} =  \widetilde{D}_{x^k}(w^s)$.  This yields the over-determined system
of {\sc pde}s 
\begin{equation}
w^a_{x^k} = T^a_k(x^i,u^\alpha_I,w^b)
\label{WE_prolongation_eqns}
\end{equation}
which is compatible if 
  $(x^i,u^\alpha_I) \in \E$. System \eqref{WE_prolongation_eqns} is referred to as the 
{\it covering equations} or the {\it Lax representation} of equation $\E$.
The variables $w^a$ are referred to as {\it pseudopotentials}.
 
The {\it evolutionary vector field} associated to an arbitrary vector-valued smooth function
$\varphi \colon J^\infty(\pi) \rightarrow \mathbb{R}^m $ is the vector field
\[
\mathbb{E}_{\varphi} = \sum \limits_{\# I \ge 0} \sum \limits_{\alpha = 1}^m
D_I(\varphi^\alpha)\,\frac{\partial}{\partial u^\alpha_I}
\]
with $D_I=D_{(i_1,\dots\,i_n)} =D^{i_1}_{x^1} \circ \dots \circ D^{i_n}_{x^n}$.

A function $\varphi \colon J^\infty(\pi) \rightarrow \mathbb{R}^m$ is called a {\it (generator of an
infinitesimal) symmetry} of equation $\E$ when $\mathbb{E}_{\varphi}(F) = 0$ on $\E$. The
symmetry $\varphi$ is a solution to the {\it defining system}
\begin{equation}
\ell_{\E}(\varphi) = 0,
\label{defining_eqns}
\end{equation}
where $\ell_{\E} = \ell_F \vert_{\E}$ with the linear differential operator
\begin{equation}
\ell_F = \left(\sum \limits_{\# I \ge 0}\frac{\partial F_r}{\partial u^\alpha_I}\,D_I\right).
\label{lF}
\end{equation}
In other words, a symmetry is a section of the {\it tangent covering} obtained by adding equation $\ell_F(\varphi)=0$  to 
equation $F=0$ and denoted by $\EuScript{TE}$.
The {\it symmetry algebra} $\mathrm{sym} (\E)$ of equation $\E$ is the linear space of
solutions to  (\ref{defining_eqns}) endowed with the structure of a Lie algebra over $\mathbb{R}$ by the
{\it Jacobi bracket} $\{\varphi,\psi\} = \mathbb{E}_{\varphi}(\psi) - \mathbb{E}_{\psi}(\varphi)$.
The {\it algebra of point symmetries} $\mathrm{sym}_0 (\E)$ is the Lie subalgebra of $\mathrm{sym} (\E)$
generated by the functions of the form $\varphi^\alpha =\eta^\alpha -\sum_{i=1}^{n} \xi^i\,u^{\alpha}_{x^i}$ with
$\eta^\alpha, \xi^i \in C^{\infty}(J^0(\pi))$.

A {\it horizontal differential k-form} on $\E$ is  the expression
\[
\omega = \sum \limits_{1\le i_1 < \dots <i_k\le n} a_{i_1 \dots i_k} dx^{i_1} \wedge \dots \wedge dx^{i_k},
\]
where  $a_{i_1\dots \i_k} \in C^{\infty}(\E)$.    The $\mathbb{R}$-linear space of such forms is denoted by 
$\Lambda^{k}_{h}(\E)$.   The {\it horizontal de Rham differential} 
$d_h \colon   \Lambda^{k}_{h}(\E) \rightarrow \Lambda^{k+1}_{h}(\E)$ acts on $\omega$ by the formula 
\[
d_h\omega = \sum \limits_{1\le i_1 < \dots <i_k\le n} \sum \limits_{l=1}^{n} D_l(a_{i_1 \dots i_k})\,dx^l  \wedge dx^{i_1} \wedge \dots \wedge dx^{i_k}.
\]
Forms such that $d_h \omega =0$ are called {\it closed}, forms $d_h\theta$ are {\it exact}. Any exact form is closed due to 
the identity $d_h \circ d_h =0$.

A {\it conservation law} on $\E$ is the cohomology class of a closed form $\omega \in \Lambda^{n-1}_{h}(\E)$. 
  A conservation law 
is said to be {\it trivial} if $\omega$ is exact, otherwise it is {\it nontrivial}. Thus, any conservation law has a representative 
\[
\omega = \sum \limits_{i=1}^{n}  (-1)^{i+1}\,a_i \,dx^{1} \wedge \dots \wedge \widehat{dx^i} \wedge \dots \wedge dx^{n}
\]
(the hat marks that the factor is omitted) such that 
\[
d_h \omega = \left( \sum_{i=1}^{n} D_i(a_i) \right) \,dx^1 \wedge \dots \wedge dx^n=0.
\]
One can write $\omega = a \,\lrcorner\, dx^1 \wedge \dots \wedge dx^n$
with $a =\sum_{i=1}^{n} a_i \,\partial_{x^i}$.

Let $\omega$ be a representative of a conservation law on $\E$, $\tilde{\omega}$ be an arbitrary extension of the form 
$\omega$ to the ambient space $J^{\infty}(\pi)$, and let $d_h\tilde{\omega}  =  \Upsilon (F) \,dx^1 \wedge \dots \wedge dx^n$, where $\Upsilon$ is  a certain differential operator in total derivatives, and $\Upsilon^{*}$ be the  adjoint operator to $\Upsilon$.  Then the function $\psi_\omega = \Upsilon^{*}(1)\vert_{F=0}$ on $\E$ is called the {\it generating function} of the conservation law.  
 
The generating function is a solution to the equation
\begin{equation}
\ell_{\E}^{*} (\psi) = \ell_{F}^{*}\vert_{F=0} (\psi) =0,
\label{cosymmetry_eq}
\end{equation} 
where $\ell_{F}^{*}$ is the adjoint operator to \eqref{lF}.  Solutions to these equations are called {\it cosymmetries} of equation 
$\E$ and their $\mathbb{R}$-linear space is denoted by $\mathrm{cosym}(\E)$. In other words, cosymmetries are sections of the 
{\it cotangent covering} obtained by appending equation \eqref{cosymmetry_eq} to $\E$  and denoted by $\EuScript{T^{*}E}$.

In general, not every cosymmetry is a generating  function of a conservation law.

In Section 5 we will use the following construction: 
 the canonical conservation law of on the Whitney sum of the tangent and the cotangent coverings, cf.
\cite[\S~6.1]{KrasilshchikVerbovetskyVitolo2017}, 
is the cohomology class  of horizontal differential $(n-1)$-forms 
defined on  $\EuScript{TE} \oplus\EuScript{T^{*}E}$  
with a representative  
$\omega$ such that   there holds  
\begin{equation}
d_h\omega = \left(\ell_F(q)\,p -q\,\ell^{*}_F(p)\right)\,dx^1 \wedge \dots \wedge dx^n.
\label{ncl_definition} 
\end{equation} 
In some cases, it is convenient to treat the variables $q_i$ and $p_j$ as odd (anticommuting), see \cite{KrasilshchikVerbovetskyVitolo2017} and references therein. Taking this into account, in \eqref{ncl_definition} and all formulas below, all instances of $q_i$ and its derivatives are placed to the left of $p_j$ and its derivatives.

Consider operator $\widetilde{\ell}_{F}$ obtained by replacing $D_{x^k}$ to
$\widetilde{D}_{x^k}$ in \eqref{lF}.
Solutions $\phi=\phi(x^i,u^\alpha_I,w^j)$ to equation $\tilde{\ell}_F(\phi)\vert_{F} =0$
are referred to as {\it shadows of symmetries in the covering} $\varpi$.
Likewise, the {\it shadows of cosymmetries in the covering} $\varpi$
are solutions to equation $\tilde{\ell^{*}}_F(\psi)\vert_{F} =0$.

\vskip 10 pt 
For the vector fields
$\pmb{a} =\sum_{i=1}^{3} a_i\,\partial_{x^i}$,  $\pmb{b} =\sum_{i=1}^{3} b_i\,\partial_{x^i}$, and the function $G$
on $\mathbb{R}^3$ we use the standard notation 
\[
\crl\, \pmb{a} = (a_{3,x^2}-a_{2,x^3}) \,\partial_{x^1} 
+ (a_{1,x^3}-a_{3,x^1}) \,\partial_{x^2}+(a_{2,x^1}-a_{1,x^2}) \,\partial_{x^3},
\]
\[
\pmb{a} \times \pmb{b}
= (a_2 \,b_3 -a_3\,b_2)\,\partial_{x^1}+(a_3 \,b_1 -a_1\,b_3)\,\partial_{x^2}
+(a_1 \,b_2 -a_2\,b_1)\,\partial_{x^3}, 
\]
\[
\mathrm{div}\, \pmb{a} =  
\sum \limits_{i=1}^{3} a_{i,x^i}, 
\qquad
\langle \pmb{a}, \pmb{b} \rangle = \sum \limits_{i=1}^{3} a_i\,b_i,
\qquad
\mathrm{grad}\,G = \sum \limits_{i=1}^{3} G_{x^i}\,\partial_{x^i}. 
\]
The expression  $\pmb{a}_{x^k}$  is understood component-wise: 
$\pmb{a}_{x^k} =\sum_{i=1}^{3} a_{i,x^k}\,\partial_{x^i}$.


\section{The three-dimensional Euler--Helmholtz equation and  its Lax representation}

Let $M \subseteq \mathbb{R}^3$ be an open star-shaped subset with local coordinates $(x^1,x^2,x^3) = (x,y,z)$,
 let $\mathfrak{vect}(M)$ be  the Lie algebra of the vector fields on $M$, and let
$\mathfrak{svect}(M) = \{\pmb{v} \in \mathfrak{vect}(M) \,\,\vert\,\, L_{\pmb{v}} \mu = 0\}$ be the Lie algebra of the volume-preserving vector fields on $M$ with the volume element $\mu = dx \wedge dy \wedge dz$.
The Lie algebra of outer derivations of $\mathfrak{svect}(M)$ is one-dimensional, \cite{Lichnerowicz1974,Morimoto1976};
we can take the vector field $\ee = x\,\partial_x+y\,\partial_y +z\,\partial_z$ as its generator. 
Let $\mathfrak{g} = C^\infty(\mathbb{R}, \mathfrak{vect}(M))$ denote the current Lie algebra of smooth 
$\mathfrak{vect}(M)$-valued functions on $\mathbb{R}$, equipped with the pointwise bracket.

The 3D Euler--Helmholtz equation, \cite[Ch. I, \S 5]{ArnoldKhesin}, reads
\begin{equation}
\crl\,\pmb{U}_t-[\pmb{U}, \crl\,\pmb{U}] =0 ,
\label{3dEH}
\end{equation}
where $\pmb{U} = U\,\partial_x+V\,\partial_y +W\,\partial_z \in  C^\infty(\mathbb{R}, \mathfrak{svect}(M))$. 
This condition yields  
\begin{equation}
\mathrm{div}\,\pmb{U} = U_x +V_y +W_z =0. 
\label{div_u}
\end{equation}
Equation \eqref{div_u} implies the compatibility of the over-determined system \eqref{3dEH}.

Since the domain $M$ is star-shaped, there exists $\uu \in \mathfrak{g}$ such that 
$\pmb{U} = \crl \uu$, see, e.g., \cite{Spivak_1}.  In terms of  the vector field $\uu$,   equation
\eqref{div_u} holds identically, while equation \eqref{3dEH} takes the form
\begin{equation}
\ccrl \uu_t-[\crl \uu, \ccrl \uu]=0,
\label{main_eq}
\end{equation}
where $\ccrl \uu = \crl (\crl \uu)$.   
Let $\pmb{F}$ denote the left-hand side of \eqref{main_eq}.  Since  $\mathrm{div}\,\pmb{F} = 0$,  equation \eqref{main_eq} is not normal, \cite[p. 7]{KrasilshchikVerbovetskyVitolo2017}.

We state the  main result of the paper:

\vskip 10 pt
\noindent
\thm  
{\it
 System \eqref{main_eq} admits the Lax representation
\begin{equation}
\left\{
\begin{array}{rcl}
\,\crl \pmb{s}_t &=& [\crl \uu, \crl \pmb{s}]+\varepsilon \, [\ee, \crl \uu],
\\
\,[\ccrl \uu, \crl \pmb{s}] &=& \lambda\, \ccrl \uu - \varepsilon \,[\ee, \ccrl \uu]
\end{array}
\right.
\label{the_covering}
\end{equation}
with $\pmb{s} \in \mathfrak{g}$.
}

\vskip 5 pt

\noindent
{\sc Proof}. Consider two derivations 
$\EuScript{D}_1 = D_t -\mathrm{ad}_{\crl \uu}$ 
and 
$\EuScript{D}_2 = \mathrm{ad}_{\ccrl\,\uu}$
of the Lie algebra $\mathfrak{g}$. 
System \eqref{the_covering} gets the form 
$\EuScript{D}_1(\crl \pmb{s})=\pmb{R}_1$, $\EuScript{D}_2(\crl \pmb{s})=\pmb{R}_2$ 
with  
$\pmb{R}_1 = \varepsilon \, [\ee, \crl \uu]$ 
and 
$\pmb{R}_2 = \lambda\, \ccrl \uu - \varepsilon \,[\ee, \ccrl \uu]$.
One readily verifies that  
$\,[\EuScript{D}_1, \EuScript{D}_2] = \mathrm{ad}_{\pmb{F}}$ and  
$\EuScript{D}_1(\pmb{R}_2) -\EuScript{D}_2(\pmb{R}_1)= \lambda \,\pmb{F} -\varepsilon \,[\ee, \pmb{F}]$.
Therefore, the compatibility conditions of system \eqref{the_covering}
are equivalent to  equation \eqref{main_eq}.
\hfill $\Box$


\section{Symmetries and local conservation laws}

The tangent covering over equation \eqref{main_eq} is obtained by appending the system
\begin{equation}
\ell_{\pmb{F}}(\qq) = \ccrl \qq_t-[\crl \qq, \ccrl \uu]-[\crl \uu, \ccrl \qq] =0.
\label{tangent_covering}
\end{equation}
Since  $\crl \qq =0$ for $\qq = \mathrm{grad}\, G$ with an arbitrary smooth function $G$ of $t,x,y,z$, all such $\qq$ are symmetries 
of \eqref{main_eq}. We will consider these symmetries as trivial.        
From straightforward computations we obtain
\vskip 10 pt
\noindent
\thm
{\it
The  quotient algebra $\mathfrak{s}$  of  the Lie algebra $\mathrm{sym}_0(\E)$ of the infinitesimal point symmetries of equation \eqref{main_eq} 
by the ideal of the trivial symmetries $\mathrm{grad} \,G$  has the following generators:
\[
\varphi_{1,0}= \ee(\uu) -2\, \uu,
\]
\[
\varphi_{1,i} = [\partial_{x^i} \times \ee, \uu],   
\]
\[
\varphi_{2,0} =t\,\uu_t+\uu, 
\]
\[
\varphi_{2,1} = \uu_t,
\]
\[
\Phi_i(A_i) = A_i\,\uu_{x^i}+A_i^{\prime}\,\left(x^i\,\partial_{x^i} +\textfrac{1}{2}\,\partial_{x^i} \times \ee\right),
\]
where  $i \in \{1,2,3\}$, 
 and $A_i=A_i(t)$  are arbitrary smooth functions. 
}
\hfill $\Box$

\vskip 10 pt
The nonzero commutators of $\mathfrak{s}$  are as follows:
\[
\,[\varphi_{1,i},    \varphi_{1,j} ] = \sum \limits_{k=1}^{3}    \epsilon_{ijk} \,\varphi_{1,k},   
\]
\[
\,[\varphi_{2,0}, \varphi_{2,1} ] = \varphi_{2,1},  
\]
\[
\,[\varphi_{1,0}, \Phi_{i}(A_i) ] = \Phi_i(A_i), 
\]
\[
\,[\varphi_{1,i}, \Phi_{j}(A_j) ] = \sum \limits_{k=1}^{3}   \epsilon_{ijk} \Phi_k(A_j), 
\]
\[
\,[\varphi_{2,0}, \Phi_{i}(A_i) ] = -\Phi_i(t\,A_i^{\prime}),  
\]
\[
\,[\varphi_{2,1}, \Phi_{i}(A_i) ] = -\Phi_i(A_i^{\prime}),   
\]
\[
\,[\Phi_{i}(A_i), \Phi_{j}(A_j) ] = 0,
\]
where $i, j, k  \in \{1,2,3\}$ and $\epsilon_{ijk}$ is the Levi--Civita symbol. 

The form of these commutators implies the following structure of the quotient algebra:  
$\mathfrak{s}=\left(\mathfrak{h}_1 \oplus \mathfrak{so}(3)  \oplus \mathfrak{h}_2 \right) 
\ltimes \mathfrak{h}_{\infty}$,
where $\mathfrak{h}_1 = \langle \varphi_{1,0}\rangle$,
$\mathfrak{so}(3) =\langle \varphi_{1,1}, \varphi_{1,2}, \varphi_{1,3}\rangle$,
$\mathfrak{h}_2=\langle \varphi_{2,0}, \varphi_{2,1}\rangle$,
and the infinite-dimensional ideal 
$\mathfrak{h}_{\infty} =\langle \Phi_{1}(A_1), \Phi_{2}(A_2), \Phi_{3}(A_3)\rangle$ is  abelian modulo trivial symmetries.

\vskip 10 pt
The cotangent covering is given by equation \eqref{main_eq} and 
\begin{equation}
\ell^{*}_{\pmb{F}}(\pp) = 
-\crl \left(\crl \pp_t -[\crl \uu, \crl \pp]\right)-[\ccrl \uu, \crl \pp]=0.
\label{cotangent_covering}
\end{equation}
We consider the  cosymmetries of the form $\mathrm{grad}\, G$ as trivial.  Then we have 
the following description:

\vskip 10 pt
\noindent
\thm
{\it 
The $\mathbb{R}$-linear space of the cosymmetries  that depend on $t, x, y, z, u,v, w$ has the following 
basis   modulo the subspace of trivial cosymmetries: 
\[
\psi_0 = \uu,
\quad \psi_i = \left(\sum_{k=1}^3 (x^k)^2 - (x^i)^2\right)\,\partial_{x^i},
\quad
\Psi_i(B_i) =B_i\,\partial_{x^i} \times \ee,
\]
where  $i \in \{1,2,3\}$ and  $B_i=B_i(t)$ are arbitrary smooth functions.
}

\hfill $\Box$

\vskip 10 pt
The conservation laws associated with these cosymmetries can be represented by the following horizontal 3-forms:
\[\fl
\omega_{\psi_0} = \Big( 
(\langle \uu, \ccrl \uu\rangle - \textfrac{1}{2}\,\langle \crl \uu, \crl \uu\rangle) \,\partial_t
- \uu_t \times \crl \uu
\]
\[
\fl
\quad
+ \uu \times (\crl \uu \times \ccrl \uu) 
\Big)\,\lrcorner\, dt \wedge dx \wedge dy \wedge dz,
\]
\[\fl
\omega_{\psi_i} = \Big(2\,\langle \ee \times \partial_{x^i}, \crl \uu\rangle \,\partial_t
-\psi_i \times (\crl \uu_t +\crl \uu \times \ccrl \uu)
\]
\[\fl\quad
-\langle \crl \uu, \crl \uu \rangle \,\ee \times \partial_{x^i}
-2\,((\ee \times \partial_{x^i}) \times \crl \uu)\times \crl \uu 
\Big) \,\lrcorner\, dt \wedge dx \wedge dy \wedge dz,
\]
\[
\fl
\omega_{\Psi_i(B)} =\Big(
\langle \Psi_i(B), \ccrl \uu\rangle \,\partial_t
- \Psi_i(B) \times (\crl \uu \times \ccrl \uu)
\]
\[
\fl\quad
+B\,\langle \crl \uu, \crl \uu\rangle \,\partial_{x^i}
-2\,B\,\langle \partial_{x^i},  \crl \uu \rangle \,\crl \uu
\]
\[
\fl\quad
+\Psi_i(B^{\prime}) \times \crl \uu
+2\,B^{\prime}\,\partial_{x^i}  \times \uu \Big) \,
\lrcorner\, dt \wedge dx \wedge dy \wedge dz. 
\]
System \eqref{main_eq}  is not normal, so the fact that the characteristics are nonzero does not guarantee the nontriviality of the corresponding conservation laws. Nevertheless, the careful analysis shows that all these conservation laws are nontrivial.


\section{Nonlocal conservation laws}

In addition to the local conservation laws $\omega_{\psi_0}$, $\omega_{\psi_i}$, $\omega_{\Psi_i(B)}$  we can derive nonlocal conservation laws using the construction of the canonical conservation law on the Whitney sum of the tangent and the cotangent coverings. 

\vskip 10 pt
\noindent
\thm
{\it For equation \eqref{main_eq} we have the canonical conservation law with the representative
\[
\fl
\Omega (\qq,\pp) = \Big( \langle \crl \qq, \crl \pp \rangle \, \partial_t +
(\crl \qq_t +\crl \uu \times \ccrl \qq-\ccrl \uu \times \crl \qq) \times \pp 
\]
\[
\qquad
+ \qq \times 
(-\crl \pp_t +[\crl \uu ,\crl \pp] +\ccrl \uu \times \crl \pp) 
\]
\[
\qquad
-\crl \qq \times (\crl \uu \times \crl \pp) \Big) \lrcorner \,dt \wedge dx \wedge dy \wedge dz
\]
}
\vskip 5 pt
\noindent 
Proof. The straightforward computation gives 
\[
d_h\Omega(\qq,\pp) = \left(\langle \ell_{\pmb{F}}(\qq), \pp \rangle - \langle \qq, \ell_{\pmb{F}}^{*}(\pp)\rangle\right)\,
dt\wedge dx \wedge dy \wedge dz.
\]
\hfill $\Box$

\vskip 10 pt
We have   the following result:  
\vskip 10 pt
\noindent
\thm
{\it
Let $\pmb{s}_{\varepsilon,\lambda}$ be a solution to system \eqref{the_covering}.  Then 
\begin{equation}
\pmb{s}_{\varepsilon,\lambda}-(\varepsilon+\lambda)\,t\,\uu
\label{the_shadow}
\end{equation}
 is a shadow of a cosymmetry in the covering 
\eqref{the_covering}, that is, a solution to system \eqref{cotangent_covering}.
}
\vskip 5 pt
\noindent
Proof.  This statement follows immediately from the easily verified identity 
\begin{equation}
\ell_{\pmb{F}}^{*}(\pmb{s}_{\varepsilon,\lambda}+\gamma\,t\,\uu) = -(\varepsilon +\lambda +\gamma)\,\ccrl \uu. 
\label{shadow_identity}
\end{equation}
\hfill $\Box$

We substitute the symmetries from $\mathfrak{s}$ instead of $\qq$ and the shadow 
\eqref{the_shadow}   instead of $\pp$ into $\Omega(\qq, \pp)$ and obtain the 3-forms 
$\Omega(\varphi_{1,0}, \pmb{s}_{\varepsilon,\lambda})$,  
$\Omega(\varphi_{1,i}, \pmb{s}_{\varepsilon,\lambda})$,  
$\Omega(\varphi_{2,k}, \pmb{s}_{\varepsilon,\lambda})$, and 
$\Omega(\Phi_{i}(A_i), \pmb{s}_{\varepsilon,\lambda})$  with $i \in \{1,2,3\}$ and $k \in \{1,2\}$.
The system \eqref{main_eq}, \eqref{the_covering} is not normal.  Appending the covering equations produce the additional obstructions to nontriviality of nonlocal conservation laws. The careful analysis show that the conservation laws  
$\Omega(\varphi_{1,i}, \pmb{s}_{\varepsilon,\lambda})$ are reducible to the local conservation laws 
$\frac{1}{2} \,(\lambda+6\,\varepsilon)\,\omega_{\psi_i}$  and   
$\Omega(\Phi_{i}(A_i), \pmb{s}_{\varepsilon,\lambda})$ are reducible to 
$\omega_{\Psi_i(K_i)}$ with $K_i = -\frac{1}{2}\,\left((\lambda+\varepsilon) \,t\,A_i^{\prime} -(\lambda+5\,\varepsilon)\,A_i\right)$,
while $\Omega(\varphi_{1,0}, \pmb{s}_{\varepsilon,\lambda})$ and $\Omega(\varphi_{2,k}, \pmb{s}_{\varepsilon,\lambda})$ 
are nontrivial and genuine nonlocal.


\section{The B{\"a}cklund transformation between the tangent and cotangent coverings}

From the form of equations \eqref{tangent_covering}  and \eqref{cotangent_covering}  
we  obtain  the following statement: 

\vskip 10 pt

\noindent
\thm
{\it System 
\begin{equation}
\left\{
\begin{array}{rcl}
\,\crl \pp_t &=& [\crl \uu, \crl \pp]+\crl \qq,
\\
\,[\ccrl \uu, \crl \pp] &=& - \ccrl \qq
\end{array}
\right.
\label{BT}
\end{equation}
defines a B{\"a}cklund transformation
$\EuScript{B}(\pp) = \qq$
 between the tangent covering  \eqref{tangent_covering}  and the cotangent covering \eqref{cotangent_covering}. 
}
\hfill $\Box$
\vskip 10 pt 
We note that system \eqref{BT} determines $\pp$ modulo the addition of an arbitrary solution $\pmb{s}_{0,0}$ to system \eqref{the_covering}  with $\varepsilon=\lambda=0$.  Taking this remark into account, we can describe the action of transformation \eqref{BT} on the cosymmetries from Theorem 3 as follows:
\[
\EuScript{B}(\psi_0)=\varphi_{2,1},
\]
\[
\EuScript{B}(\psi_i)=-2\,\varphi_{1,i}, 
\]
\[
\EuScript{B}(\Psi_i(B_i))=2\,\Phi_{i}(B_i).
\]
 These formulas also describe the action of the inverse transformation $\EuScript{B}^{-1}$  on the symmetries $\varphi_{2,1}$, $\varphi_{1,i}$, and $\Phi_{i}(A_i)$, while the action of $\EuScript{B}^{-1}$ on $\varphi_{1,0}$ 
and $\varphi_{2,0}$ is given in terms of the pseudopotential $\pmb{s}_{\varepsilon,\lambda}$:
\vskip 10 pt
\noindent
\thm
{\it     We have 
$\EuScript{B}^{-1}(\varphi_{1,0})=\pmb{s}_{1,-1}$ and $\EuScript{B}^{-1}(\varphi_{2,0})=\pmb{s}_{0,-1} +t\,\uu$.
}
\vskip 5 pt 
\noindent
Proof.  This  follows from  \eqref{shadow_identity} and the equality
$\EuScript{B}(\pmb{s}_{\varepsilon,\lambda}+\gamma\,t\,\uu) = \varepsilon \, \varphi_{1,0} +\gamma\, \varphi_{2,0}$.

\hfill $\Box$


\section*{Acknowledgments}

I would like to express my sincere gratitude  to I.S. Kra\-{}sil${}^{\prime}$\-{}shchik for  very important discussions.

Computations  were done using the {\sc Jets} software \cite{Jets}.


\bibliographystyle{amsplain}

\end{document}